\documentclass[11pt]{article}
\usepackage{mathrsfs}
\usepackage{epic, eepic}
\usepackage{subfigure}
\usepackage{amsfonts, amssymb, amsmath, graphicx, epsfig, lscape, capt-of, xcolor}
\usepackage{url}
\usepackage{ifpdf}
\def\inst#1{$^{#1}$}

\makeatletter

\begin{document}%

%%%%%%%%%%%%%%%%%%%%%%

\title{Measuring network resilience through connection patterns}

\author{%
Roy Cerqueti\inst{1}\and
Giovanna Ferraro\inst{2} \and
Antonio Iovanella\inst{2}\thanks{Corresponding author.}}
\date{}

\maketitle

\begin{center}
{\footnotesize
\inst{1} Department of Economics and Law\\
University of Macerata\\
Via Crescimbeni, 20 - 62100 Macerata, Italy\\
\texttt{roy.cerqueti@unimc.it}\\
\vspace{0.3cm} \inst{2} Department of Enterprise Engineering\\
University of Rome Tor Vergata\\
Via del Politecnico, 1 - 00133 Rome, Italy.\\
\texttt{giovanna.ferraro@uniroma2.it\\
antonio.iovanella@uniroma2.it}}
\end{center}

\begin{abstract}
Networks are at the core of modeling many engineering contexts,
mainly in the case of infrastructures and communication systems. The
resilience of a network, which is the property of the system capable of absorbing 
external shocks, is then of paramount relevance in the
applications. This paper deals with this topic by advancing a
theoretical proposal for measuring the resilience of a network. The
proposal is based on the study of the shocks propagation along the
patterns of connections among nodes.
The theoretical model is tested on the real-world instances of two
important airport systems in the US air traffic network; 
Illinois (including the hub of Chicago) and New York states (with JFK
airport).

\vspace{5 mm} Keywords: networks; resilience; paths; weighted arcs;
air traffic systems.
\end{abstract}

%%%%%%%%%%%%%%%%%%%%%%%%%
%%%%%%%%%%%%%%%%%%%%%%%%%
\section{Introduction}

Networks are an important tool used to describe and analyze the
structure and dynamical behaviors of several complex systems found
in the real world. In particular, network representation applies to
transportation systems such as airline routes \cite {CBBV} and
railroads \cite{SFLT}; large communication infrastructures such as
the World-Wide-Web  \cite {BAJ}
and the Internet \cite {FFF};
biological systems \cite {Alon} such as gene and protein interaction
networks \cite  {RVV} and many social interaction structures
\cite {BMBL}, \cite {Faust}.

In the framework of networks, the interacting elements are depicted
as nodes in the system and their interactions are represented as links connecting
the nodes. A relevant issue relies to the behavior of
networks with respect to the occurrence of external shocks.

In this respect, it is possible to distinguish the network's {\it
robustness}, i.e. the capacity of the network to perform its basic
functions even in the event of missing elements (nodes or links) or, to phrase this concept
differently, the ability of the system to tolerate strains and to
maintain its functionality in some way; and the network's {\it resilience}, i.e. the ability of a system to return to its
normal state after a disturbance due to internal or
external shocks by altering its processes while continuing to
perform \cite{barbook}, \cite{FI3}.

%, the so-called \emph{resilience} of the network. %The
%analysis of tolerance against external shocks as for example random
%failures or intentional attacks allows us to consider whether such
%shocks can occur.

Literature on networks resilience is rather wide. At its core, shocks
are interpreted as the removal of nodes on the basis of several different
criteria (see for example: \cite {AJBerror}, \cite {CC}, \cite {CFI2}, \cite {FI3}).

%Ceptureanu ????

However, engineering applications reveal that shocks can not be
associated forcefully to the disappearance of nodes. One might
experience the occurrence of a shock as a perturbations whose size
proxies the shock's relevance. An example can be the loss
of capacity of an electricity plant due to adverse weather
conditions and the corresponding damage of the entire energy system.

The network resilience is considered crucial for the different
practical approaches whereby the criticalities related to the
eventual failure of nodes and links and by means of overall system
tolerance. Disruptive events, whether a natural disaster, intentional
attacks or common failures, can have relevant impacts  when they
lead to the failure of network elements. Indeed, network
performances are directly related to their resilience and thus to
the abilities of networks in tolerating loss of important elements
such as bridges or hubs \cite{CFI2}.

Resilience has been studied across several different network structures
and there is now knowledge regarding how specific types of networks react to specific kinds of losses \cite{BVFHH}.
Studies about this topic have considered a large variety of structural failures, both induced by attack
or naturally occurring, which involve nodes chosen by their central position in the system, as well as those that are random.

However, the current literature on how to measure resilience is very
limited \cite{BRMR}; \cite{HBRM}; \cite{Woods}
with most contributions focusing on the remaining
part of the network after the damage. For example, some authors
measured resilience through the number of node failures and link
failures that can occur for a  system that is still functioning
\cite{Rosenkrantz}; others proposed an approach to evaluate the
logistics of network resilience based on the redundant resource and
reliability of nodes and links \cite{WI}; Ash and Newth \cite{AN}
used an algorithm to evolve the complex systems that are resilient
to cascading failure; Wang et al., \cite{WG} proposed a measure
based on the recovery ability of a system.

In this context, it is therefore relevant to analyze how an external shock that affects a node propagates
along the network paths.

The aim of this paper is to introduce a new analytical measure of resilience for weighted networks that can be
used when external shocks occur. The idea is to subject the network to something akin to a load test and check how
it responds.

Thus, we consider the resilience of a network as its ability to
absorb a shock. We introduce the concept of shock and the modality
of its propagation, which is assumed to depend strongly on the
weights of the arcs. The presence of weights along the links allows
us to take into account the relevance given to the length of the
paths between the nodes.

The intuitive basic idea is that a shock occurs at a single node and
it propagates from there over all the arcs, starting from the
involved node. Once the shock reaches the adjacents, it then
proceeds to the adjacents of the adjacents, and so on. There is a
required condition for the propagation, based on the weights on the
arcs. Substantially, the shock propagates only if the propagating
arcs have sufficient weight. Moreover, we also assume that an
amplifying/damping effect on the shock is in force on the basis of
the distance from the originating node. Details will be provided
below.

We test the proposed measure of resilience on two empirical networks
extracted from the network of US commercial airports: the Illinois state network and the
New York state network.

The paper is organized as follows. Section 2 provides preliminaries
and notations. Section 3 focuses on the conceptualization of the
proposed measure of the resilience. Section 4 shows the analysis of
the resilience on two empirical networks. Section 5 offers
concluding remarks. %with discussions of the future works.

%%%%%%%%%%%%%%%%%%%%%%%%%
%%%%%%%%%%%%%%%%%%%%%%%%%
\section{Preliminaries and notations}

A network is an analytical-geometrical structure which is able to
describe a set of interconnected elements in a unified system. The
basis of a network is a graph $G = (V, E)$, where $V$ is the set of
$n$ nodes (or vertices) and $E$ is the set of $m$ links (or arcs).
The links formalize the connections among the nodes.

Two generic nodes will be denoted by two integers $i,j=1,\dots, n$.
The presence of a link between the nodes $i$ and $j$ will be
captured through a binary variable $a_{ij}$ which is $1$ or $0$ if
the link does exist or not, respectively.

In so doing, we can fully identify $E$ through the adjacency
matrix $\mathbf{A}=(a_{ij})_{i,j=1, \dots,n}$. %  The graph is
%symmetric when $a_{ij}=a_{ji}$, for each $i,j=1, \dots, N$, and
%asymmetric otherwise. In this latter case, links are (and the graph
%is) oriented.
The proposed setting is grounded on networks which are weighted and
oriented. Specifically, two possible arcs connect $i$ and $j$: the
one from $i$ to $j$ and the one from $j$ to $i$. Accordingly, we
will hereafter denote the arc from $i$ to $j$ by $i \to j$.
Moreover, arcs are weighted by positive numbers which proxy the
level of interaction among the nodes. We denote the weight of the
arc $i \to j$ by $w_{ij}$. In general, $w_{ij} \neq w_{ji}$. By
assuming that $a_{ij}=0$ if and only if $w_{ij}=0$, we can replace
the matrix $\mathbf{A}$ with the more informative weighted adjacency
matrix $\mathbf{W}=(w_{ij})_{i,j=1, \dots,n}$. Indeed, by employing
such a matrix, one can simultaneously identify the existing links
and their corresponding weights.

The network is then $N=(V, \mathbf{W})$.

Two arcs are said to be consecutive if they share exactly one of the
two nodes compounding them. Two consecutive arcs form a path of
length 2 or, simply, a 2-path. A generic 2-path with starting node
$i_0 \in V$ can be written as $p_{i_0}^{(2)}=\{i_0 \to i_1 \to
i_2\}$, with $i_0,i_1, i_2$ being three distinct nodes in $V$ and $i_0
\to i_1,i_1 \to i_2 \in E$.

Given an integer $k \geq 3$, we can extend such a definition to the
case of $k$-paths in a natural way. A generic $k$-path with starting
node $i_0 \in V$ is $p_{i_0}^{(k)}=\{i_0 \to i_1 \to \dots \to
i_{k-1} \to i_k\}$, where $i_0,i_1,\dots, i_{k-1},i_k$ are distinct
nodes in $V$ and $i_{h-1} \to i_h \in E$, for each $h=1,\dots, k$.

Under this definition, a 1-path is simply an arc in $E$.

We collect all the $k$-paths of $N$ with starting node $i_0$ in a
set $\mathcal{P}^{(k)}(i_0)$. All the $k$-paths of the network will
be collected in
\begin{equation}\label{Pk}
\mathcal{P}^{(k)}=\bigcup_{i_0 \in V} \mathcal{P}^{(k)}(i_0).
\end{equation}

The weight of a $k$-path can be defined by the vector of the weights
of the arcs composing it. We denote such a weight by
$w(p_{i_0}^{(k)})=(w_{i_0i_1},w_{i_1i_2}, \dots,w_{i_{k-1} i_k})$. A
suitable aggregation of the components of the weight of
$p_{i_0}^{(k)}$ will be discussed in the next section, in the
context of the propagation of the shocks in the network.

We also denote by $\bar{k}$ the length of the longest $k$-path of
the network.

%%%%%%%%%%%%%%%%%%%%%%%%%
%%%%%%%%%%%%%%%%%%%%%%%%%
\section{The measure of the resilience}\label{model}

This section is devoted to the conceptualization of the measure of
the resilience of the network.

The resilience of a network describes its ability to absorb a shock.
Therefore, in order to provide a definition of resilience, one needs
to introduce the concept of shock and also the way in which it
propagates in the network.

A shock is an external solicitation of one of the nodes of the
network. Such a local occurrence is able to destabilize the entire
network by propagating over the other nodes of the network. The
destabilization of the network is strongly dependent on how shocks
propagate and, of course, on the entity of the shock. Indeed, shocks
are not identical, and they can be of different size. We 
measure the size of a shock by a scalar $\xi \in [0,+\infty) $, with
the conventional agreement that the higher the value of $\xi$, the
more severe the occurred shock. The case $\xi=0$ is included for the
sake of completeness, and it means \emph{no shock}.

In this framework, we assume that the propagation of a shock
occurring at a given node $i_0$ follows the route traced by the
$k$-paths with starting node $i_0$, and depends on the values of the
weights of the arcs of the $k$-path and on the distance of the nodes
of the $k$-path from the shocked node.

We enter the details. 

Consider an integer $k$ and a $k$-path having, as a starting node, a node with a shock,
%( ?  with starting node the shocked one ), 
namely $p_{i_0}^{(k)}=\{i_0 \to i_1 \to \dots \to
i_{k-1} \to i_k\} \in \mathcal{P}^{(k)}(i_0)$. Hypothetically, the
shock propagates on a sequential basis: from $i_0$ to $i_1$, from
$i_1$ to $i_2$ and so on, till the last node $i_k$. In real
practice, the shock propagation is dampened or amplified -- it
depends on the specific context -- as the distance from the starting
node grows. As we will see, such a damping/amplifying effect might
eradicate/exacerbate the action of the shock at nodes that are far
enough from $i_0$.

The way the shock propagates is assumed to be captured by the weight
$w(p_{i_0}^{(k)})$. This is a natural requirement, in that the
propagation of a shock from a node to another one has necessarily to
be related to the entity of the interaction between them.

According to these arguments, we introduce a discount factor $\delta
\in [0,+\infty)$, which describes the propagation motion of a shock
with size $\xi$ on $p_{i_0}^{(k)}$ in a discrete time framework, as
follows:
\begin{itemize}
\item at time $t=0$, the shock with size $\xi_0=\xi$ occurs at $i_0 \in V$;
%\item at time $t=1$, the shock propagates to $i_1$ with size $\xi w_{i_0i_1}
%\delta$;
%\item at time $t=2$, the shock propagates to $i_2$ with size $\xi [w_{i_0i_1}
%\delta+w_{i_1i_2}] \delta$;
\item at a generic time $t=h$, with $1 \leq h \leq k$, the shock propagates to $i_h$ with size $\xi_h$, with $\xi_h=[\xi_{h-1}+w_{i_{h-1}i_h}]\delta=
\xi \sum_{s=1^h}w_{i_{s-1}i_s} \delta^{h-s+1}$. If $h < k$, then the
shock propagates to node $i_{h+1}$ only if a \emph{propagation
condition} is satisfied (see below). Once the shock ceases to
propagate, the propagation motion stops.
\end{itemize}
Notice that $\delta \in (0,1)$ is associated to a damping of the
distance from the shocked node, while $\delta>1$ captures an
amplifying effect. The case $\delta=0$ is trivial, and means ``no
propagation", while $\delta=1$ means that the distance from the
original node has no impact on the propagation of the shock.

The \emph{propagation condition -- PC,} hereafter -- is a
requirement to be satisfied in order to avoid the stopping of the
propagation motion. In this model, we assume that the shock
propagates only if its size is \emph{large enough}. By recalling
that the longest $k$-paths in the network are the ones with
$k=\bar{k}$, we
introduce a vector of positive scalars $\Gamma =(\gamma_1, \dots \gamma_{\bar{k}}) \in \mathbb{R}^{\bar{k}}$.%
%such that $0<\gamma_{h-1}<\gamma_{h}$, for each $h=2,\dots, k $.

Given $h=1,2,\dots, k$, we assume that the shock does not propagate
to $i_h$ if and only if there exists $s=0,1,2,\dots,h-1$ such that
$\xi_s < \gamma_s$.

Given $\Gamma \in \mathbb{R}^{\bar{k}}$ and $\xi \in [0,+\infty)$,
if the shock is propagated till the last node $i_k$, then we say
that $p_{i_0}^{(k)}$ satisfies $(\Gamma,\xi)-PC$ or, simply,
$p_{i_0}^{(k)}$ is $(\Gamma,\xi)-PC$. Vector $\Gamma$ is the
\emph{PC-vector} of the network $N$.

Notice that the thresholds $\gamma$'s can be taken as endogenous,
and are fixed a priori. In principle, they do not depend on the
specific starting node $i_0$. Therefore, we can identify the PC of
the entire network by defining the minimal values of the $\gamma$'s
serving for the propagation condition. We denote them by
$\gamma^\star$'s and collect them in $\Gamma^\star=(\gamma^\star_1,
\dots \gamma^\star_{\bar{k}})$. Vector $\Gamma^\star$ could be
defined as the \emph{critical PC-vector} of the network $N$.

Clearly, specific values of $\Gamma$ and $\xi$ determine the set of
the $k$-paths which are $(\Gamma,\xi)-PC$. In this respect, fixed
$\Gamma \in \mathbb{R}^k$ and $\xi \in [0,+\infty)$, we define
\begin{equation}
\label{PkPC} \mathcal{P}^{(k)}_{\Gamma,\xi}=\bigcup_{i_0 \in V}
\left\{ p^{(k)}_{i_0} \in \mathcal{P}^{(k)}(i_0): p^{(k)}_{i_0}
\text{ is }(\Gamma,\xi)-PC\right\}.
\end{equation}

We can define the $\Gamma-\xi$-\emph{resilience} of the network as
the ability of the network to stop the propagation of the shocks
when the PC-vector is $\Gamma^\star$ and the shocks have initial
size $\xi_0=\xi$.

The sets defined in (\ref{Pk}) and (\ref{PkPC}) can be employed to
provide a relative measure of the resilience of $N$, once $\Gamma$
and $\xi$ are fixed. In particular, one can define the
\emph{measure} of the $\Gamma-\xi$-resilience of $N$ -- and denote
it by $\mu_{(\Gamma,\xi)}(N)$ -- as a weighted mean of the relative
number of the $k$-paths which are $(\Gamma,\xi)-PC$ as $k$ varies:
\begin{equation}
\label{mu} \mu_{(\Gamma,\xi)}(N)=\sum_{k=1}^{\bar{k}}\theta_k\frac{
|\mathcal{P}^{(k)}_{\Gamma,\xi}|}{|\mathcal{P}^{(k)}|},
\end{equation}
where the weights are collected in a vector of nonnegative scalars
$\Theta=(\theta_1, \dots, \theta_{\bar{k}})$ such that
$\sum_{k=1}^{\bar{k}}\theta_k=1$. By construction,
$\mu_{(\Gamma,\xi)}(N) \in [0,1]$, for each network $N$, $\Gamma \in
\mathbb{R}^{\bar{k}}$ and $\xi \in [0,+\infty)$. Such a property
allows one to compare different networks in terms of their abilities to
absorb shocks.

Furthermore, the presence of the weights $\theta$'s allows one to
provide different conceptualizations of $\Gamma-\xi$-resilient
networks, on the basis of the relevance given to the length of the
paths.

The corner cases of $\mu_{(\Gamma,\xi)}(N)=0$ and
$\mu_{(\Gamma,\xi)}(N)=1$ stand for the minimum and maximum level of
$\Gamma-\xi$-resilience, respectively. Indeed, in the former case
the shocks with size $\xi$ are immediately absorbed; in the latter
case, any shock of size $\xi$ propagates over all the available
paths.

Intuitively, $\mu_{(\Gamma,\xi)}(N)$ decreases with respect to all
the components of $\Gamma$ and increases with respect to $\xi$,
because PC becomes more restrictive if the thresholds $\gamma$'s are
large or if the size of the shock is small.

%For a review of the employed notation, please refer to Table
%\ref{tab_2}.
%
%\begin{table}[t]
%\begin{footnotesize}
%%\begin{tiny}
%\begin{center}
%\begin{tabular}{|c|l|}
%\hline
%Symbol & Meaning\\
%\hline\hline
%$N$     & The graph $G$ and its matrix of weight $\mathbf{W}$.\\
%$\bar{k}$       & Max path length.\\
%$\xi$           & Size of a shock, $\xi \in [0,+\infty).$\\
%$\delta$        & Discount factor, $\delta \in (0,1)$.\\
%$\Gamma$    & Vector of propagation thresholds with $\Gamma=(\gamma_1, \dots \gamma_{\bar{k}})$.\\
%$\mu_{(\Gamma,\xi)}(N)$       & $\Gamma-\xi$-resilience of $N$.\\
%$\Theta$        & Vector of weights for the convex combination in $\mu_{(\Gamma,\xi)}(N)$ with $\Theta = \{\theta_1, \theta_2, \dots, \theta_{\bar{k}}\}$.\\
%\hline \hline
%\end{tabular}
%\end{center}
%\caption{Table of notation.} \label{tab_2}
%%\end{tiny}
%\end{footnotesize}
%\end{table}

%%%%%%%%%%%%%%%%%%%%%%%%%
%%%%%%%%%%%%%%%%%%%%%%%%%

\section{Applications}

Herein we considered the analysis of the resilience on two empirical
networks extracted from the network of US commercial
airports~\cite{CPV}. The original US commercial airports network has
$n = 500$ nodes denoting airports and $m = 2980$ directed arcs
representing flight connections. In this network, weights are the
number of daily flights available on such connections in 2010. The
network has both a small-world and scale-free organization
%with $\gamma \simeq 1.8$
~\cite{BBPV2}. In particular, we present the
cases of the airports in Illinois (IL) and New York (NY) states. The
selected networks are relatively small. This allows us to overcome the
computational complexity problems, since the computation of all the
simple paths grows exponentially with the size of the
network~\cite{FHW}. 
However, the case studies regarding the airports in Illinois and 
New York states can be considered of interest being that the
two airport hubs of Chicago (ORD according to International Air 
Transport Association - IATA - code nome) 
and New York (JFK) are extremely crucial in 
US air traffic network.

\begin{figure}[!htb]
    \centering
        \includegraphics[scale = 0.6]{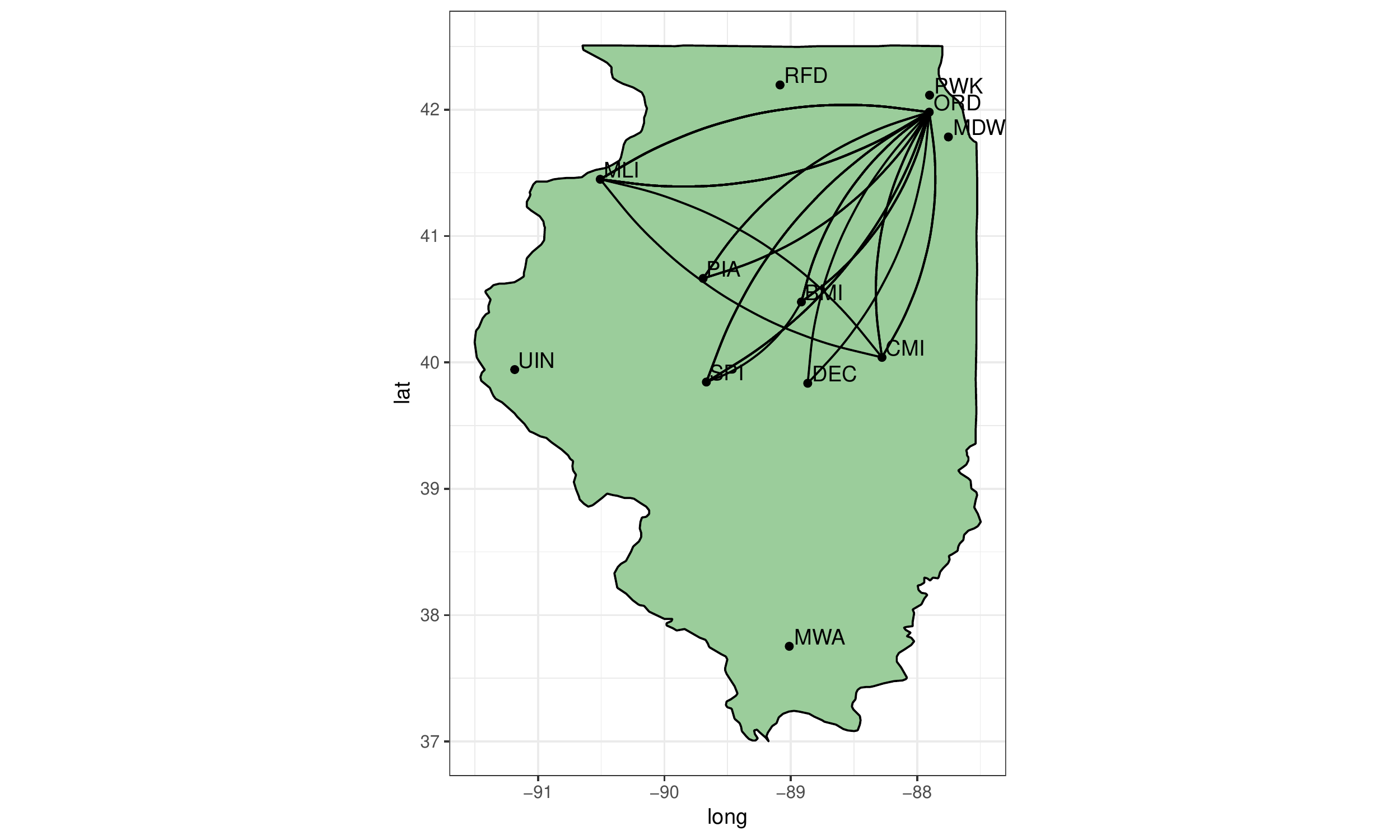}
        \caption{Map of airports and routes for Illinois state.}
        \label{mapsIL}
\end{figure}

\begin{figure}[!htb]
\centering
    \includegraphics[scale = 0.6]{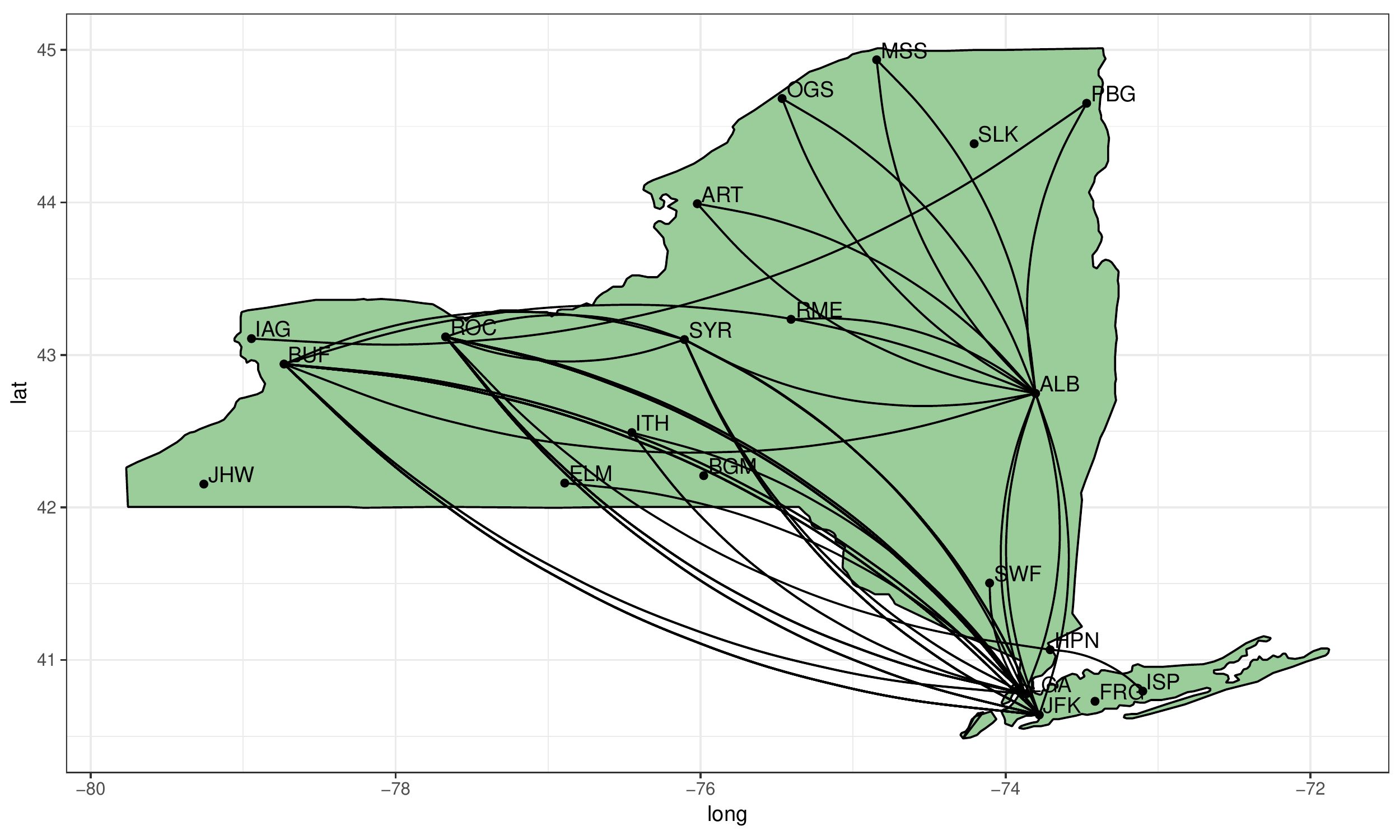}
    \caption{Map of airports and routes for New York state.}
        \label{mapsNY}
\end{figure}

The subnetwork extracted for IL is composed of $12$ nodes and $51$
arcs with $\bar{k} = 4$ while the NY subnetwork is composed of $21$
nodes and $89$ connection arcs with $\bar{k}= 8$.
Figures~\ref{mapsIL} and~\ref{mapsNY} show, respectively, the visualizations of
relative networks overlapped  on the maps of NY and IL states.

The data processing, the network analysis and all simulations are
conducted using the software R~\cite{RCT} with the {\it igraph}
package~\cite{CN}. The datasets were obtained from the R packege {\it
tnet} authored by Tore Opsahl (http://toreopsahl.com). Code in the R
programming language is available upon request.
%
%For the sake of readability we report in Table~\ref{tab_2} the
%notations used hereafter.

Formulas (\ref{PkPC}) and (\ref{mu}) were implemented in an
algorithm which takes as input the network $N$ under observation and
some different settings as follows.

We consider three different settings for $\Gamma$, in particular:
$\Gamma_1 = \{\gamma_i = 1, i = 1, ..., \bar{k}\}$, i.e. the
propagation thresholds are low and a large number of shocks
propagate; more than this, all the shocks propagate when $\delta=1$;
$\Gamma_2 =\{\gamma_i = 1, i = 1, ..., \lceil \bar{k}/2\rceil$ and
$\gamma_i = 2^{i-\lceil\bar{k}/2\rceil+1}$ for $i = \lceil \bar{k}/2
\rceil + 1,..., \bar{k}\}$, i.e shocks propagation becomes more
difficult as the distance from the shocked nodes increases; finally,
$\Gamma_3 =\{\gamma_i = 2^{\lfloor\bar{k}/2\rfloor-i+1}$ for $i =
1,.., \lfloor \bar{k}/2 \rfloor$ and $\gamma_i = 1$, $i = \lfloor
\bar{k}/2 \rfloor + 1, ..., \bar{k}\}$, i.e. the propagation is
obstructed along shortest simple paths with a decreasing geometric
evolution.

Regarding the weights in $\Theta$, we considered as first setting
$\Theta_1$ whose $i$-th component is $\theta_i = 1/\bar{k}$, for $i
= 1, \ldots, \bar{k}$; a second is $\Theta_2$ with $\theta_i =
1/2^i$ for each $i = 1, ... ,\bar{k}-1$ and $\theta_{\bar{k}} =
\theta_{\bar{k} -1}$; a third setting is $\Theta_3$ with $\theta_i =
1/2^{\bar{k}-i+1}$ for each $i = 2, ... ,\bar{k}-1$ and $\theta_{1}
= \theta_{2}$.

In the last two settings, coefficients are taken from a geometric
progression of decreasing and increasing type, respectively, while
the first one is associated to uniform values of $\theta$'s.% plus
%the last member equal to the second to last in $\Theta_2$ and first
%member equal to the second in $\Theta_3$, respectively, i.e. the
%segment $0-1$ is divided in a progression of half subsegments of
%decreasing or increasing length.

Finally, for the computation of the $\mu_{(\Gamma,\xi)}(N)$, we
considered $\xi = \{0, 1, 2, \ldots, 10\}$ and $\delta = \{0, 0.1,
0.2, \ldots, 1\}$, thus we performed 121 computations for each
network. The higher values of $\xi$ were chosen in order to be
greater than $\bar{k}$.

%\begin{figure}[!htb]
%    \centering
%    \begin{minipage}{0.5\textwidth}
%        \centering
%        \includegraphics[scale = 0.4, trim = 1cm 0cm 1cm 0cm, clip]{mappa_ny}
%        %\caption{a}
%    \end{minipage}%
%     %\ \hspace{5mm} \hspace{5mm}
%    \begin{minipage}{0.5\textwidth}
%        \centering
%        \includegraphics[scale = 0.4, trim = 2cm 0cm 1cm 0cm, clip]{mappa_il}
%        %\caption{b}
%    \end{minipage}
%    \caption{Maps of airports and routes for New York (left) and Illinois (left).}
%    \label{maps}
%\end{figure}

%%%%%%%%%%%%%%%%%%%%%%%%%
\subsection{Illinois state network}

\begin{figure}[!htb]
    \centering
    \begin{minipage}{0.3\textwidth}
        \centering
        \includegraphics[scale = 0.28, trim = 4cm 0cm 5cm 0cm, clip]{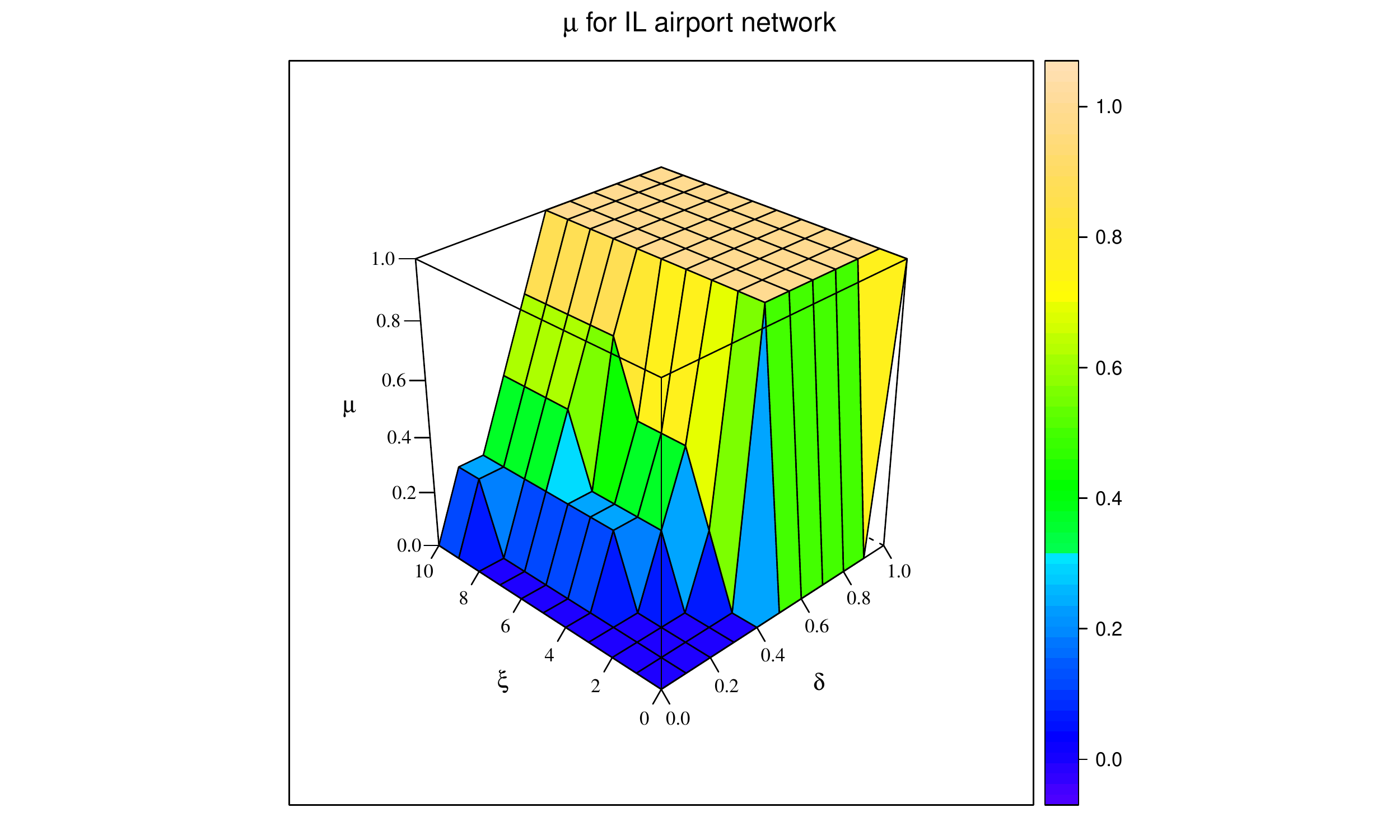}
        %\caption{$\Theta_1$}
    \end{minipage}%
     %\ \hspace{5mm} \hspace{5mm}
    \begin{minipage}{0.3\textwidth}
        \centering
        \includegraphics[scale = 0.28, trim = 4cm 0cm 5cm 0cm, clip]{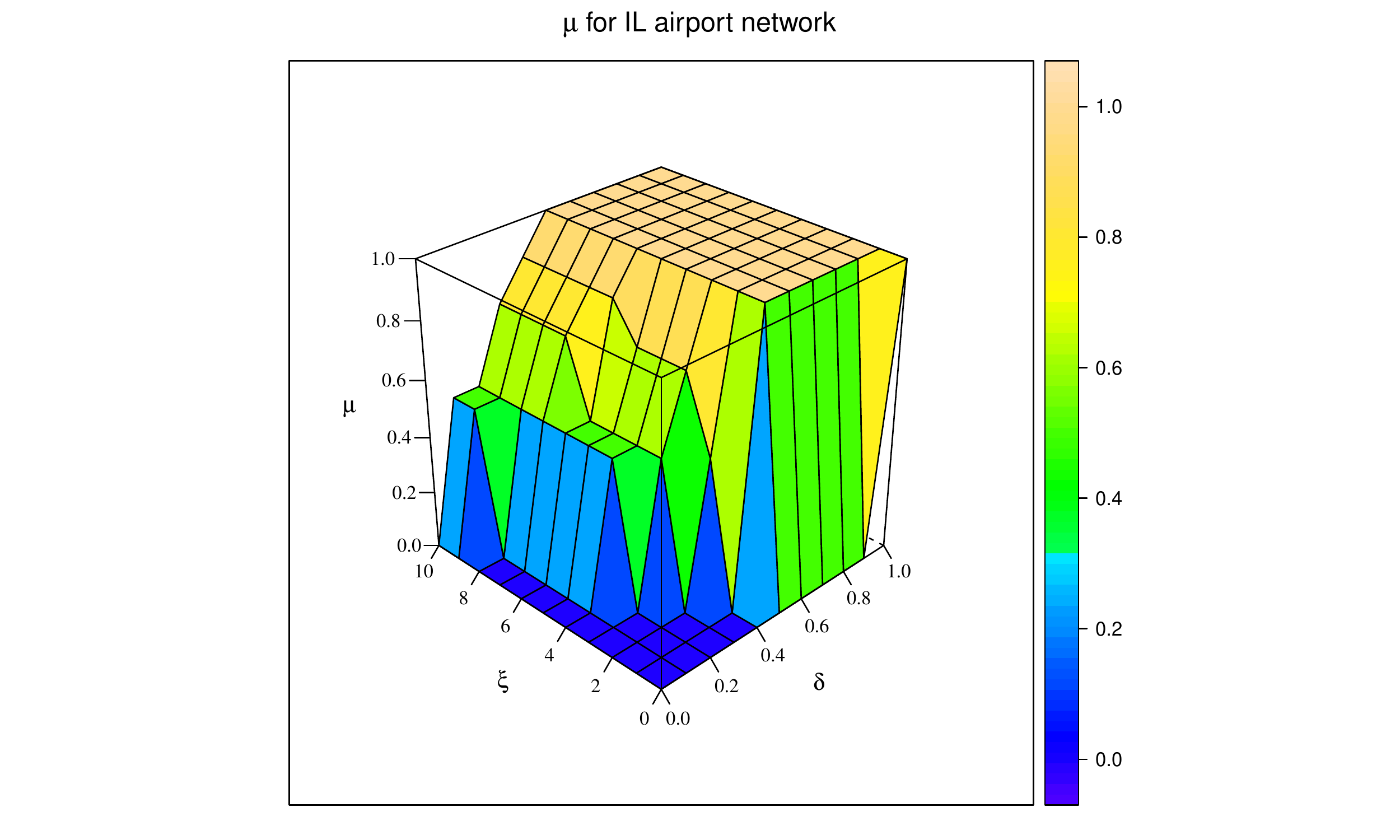}
        %\caption{$\Theta_2$}
     \end{minipage}%
     \begin{minipage}{0.3\textwidth}
        \centering
        \includegraphics[scale = 0.28, trim = 4cm 0cm 5cm 0cm, clip]{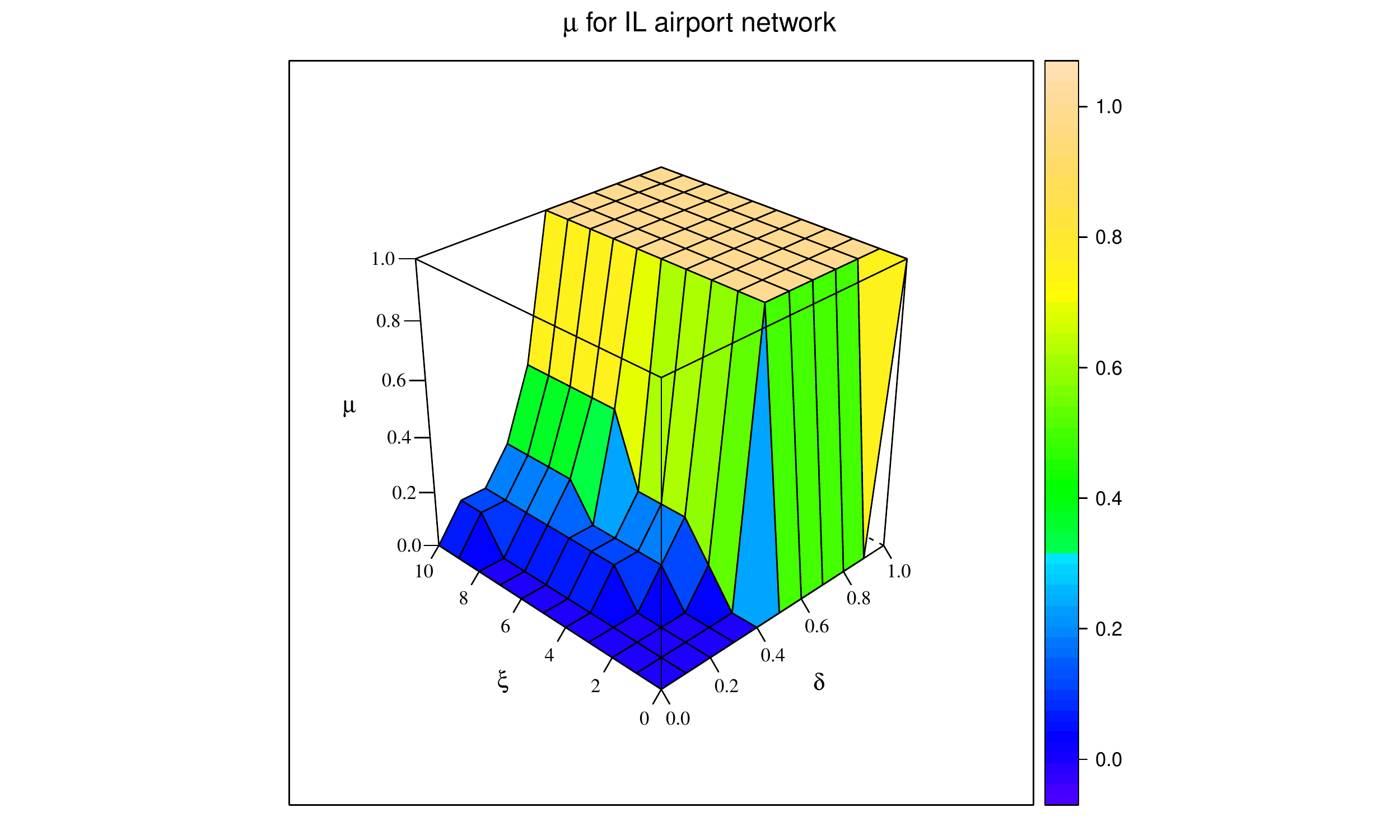}
        %\caption{$\Theta_3$}
     \end{minipage}
    \caption{$\mu_{(\Gamma,\xi)}(N)$  values in case of IL state airports for $\Gamma_1$ and $\Theta_1$ (left), $\Theta_2$ (center) and $\Theta_3$ (right).}
    \label{il_1}
\end{figure}

Figure~\ref{il_1} reports the results for the IL airports for
$\Gamma_1$ and the three different settings of $\Theta$. All the
cases share the intuitive behavior of the resilience such that for
small values of the $\xi$'s and $\delta$'s, the value of
$\mu_{(\Gamma,\xi)}(N)$ is small and increases with respect to
$\delta$ and $\xi$, becoming equals to 1 when $\delta$ and
$\xi$ are large enough. However, some remarkable differences emerge.

For cases $\Theta_1$ and $\Theta_3$, small values of the discount
factor leads to hard shock propagation, even if the size of the
shock is large. This is not longer true for $\Theta_2$, where an
enlargement of the shock is able to increase the level of the
resilience also when the discount factor is at a low level. This outcome is
due to the more prominent role played by the shortest paths in the
first and third setting when compared to
the second one, where the shortest
paths are strongly penalized by the weights. In this respect, one
should also consider that the action of the discount is more evident
as the distance with the shocked node increases.

Similar arguments can also explain the discrepancies in the rate of
growth of the resilience with respect to $\delta$. Indeed,
the increase $\mu_{(\Gamma,\xi)}(N)$ is faster for $\Theta_3$,
while the lower rate can be observed in the $\Theta_2$ case.
Specifically, for $\Theta_3$ and for each value of $\xi$, we can
identify the resilience as collapsing to a certain degree, with the network
passing rapidly from resilience equal to 1 to very low resilience
when $\delta$ moves slightly below 0.5.

The same collapse of the resilience can be observed for all the
cases when the size of the shock goes below 2. Thus, one can argue
that there exists a critical threshold for the shock leading to
unitary resilience, for each value of the discount factor. This is
totally in agreement with $\Gamma_1$, which has components with
minimum value. In the specific case of the IL airports, it is then
associated to the propagation of all the shocks large enough when
$\delta<1$ or all the shocks when $\delta=1$.

%[XXXX RIMUOVEREI, O NON SO SE SI PUO' INTEGRARE QUALCOSA is small
%then the shock should be of high value to permit propagation and
%small shocks propagate only if the discount factor is high. We
%obtain similar trend for $\Theta_2$ but with a higher slope, that
%means that we obtain the value of $\mu_{(\Gamma,\xi)}(N) = 1$ for
%the same combinations of $\xi$ and $\delta$, but for smaller value
%of the latter, $\mu_{(\Gamma,\xi)}(N)$  has higher values. It means
%that when we weight more the shorter paths, then the propagation
%tends to be more prone to happen, as it is logical to obtain. XXXX]

\begin{figure}[!htb]
    \centering
    \begin{minipage}{0.3\textwidth}
        \centering
        \includegraphics[scale = 0.28, trim = 4cm 0cm 5cm 0cm, clip]{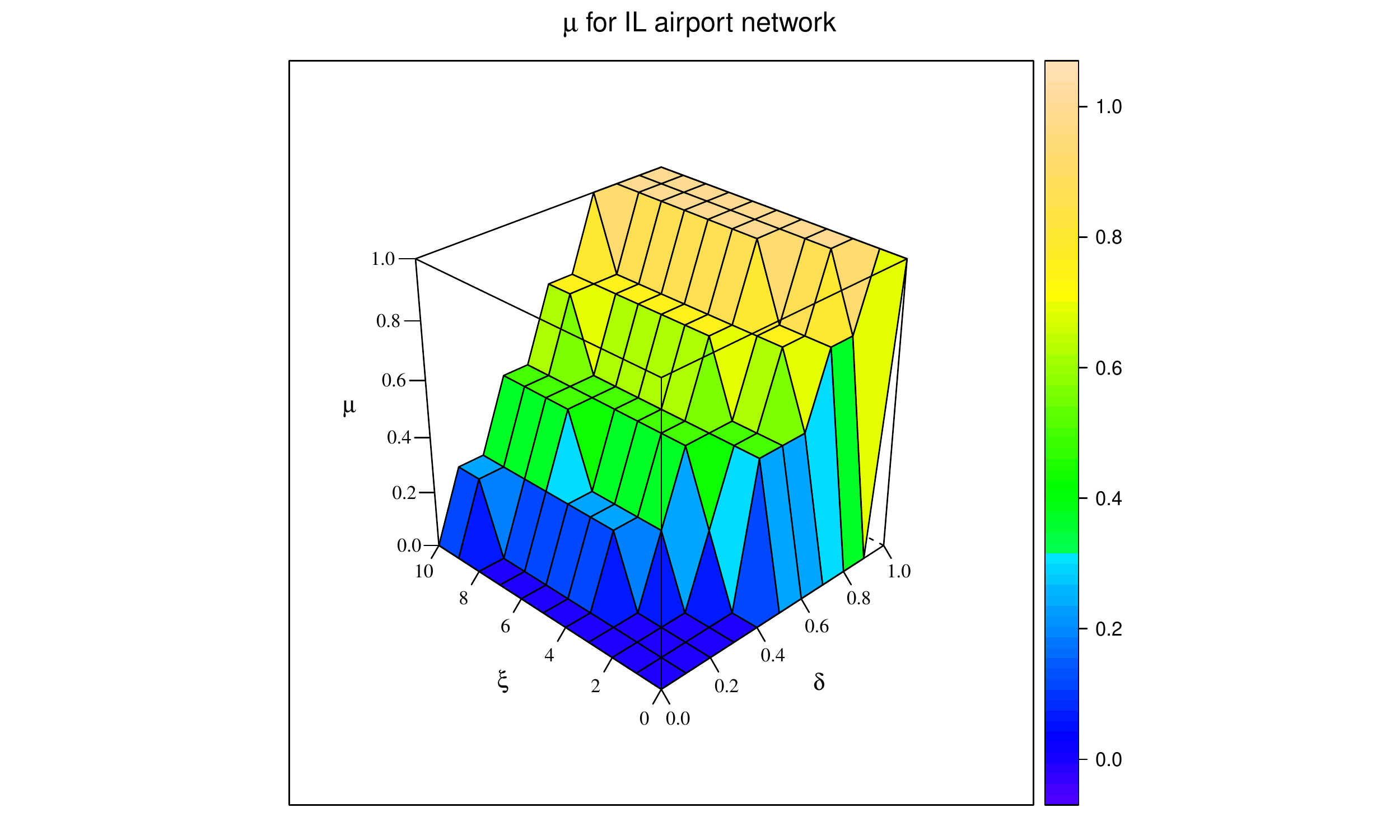}
        %\caption{$\Theta_1$}
    \end{minipage}%
     %\ \hspace{5mm} \hspace{5mm}
    \begin{minipage}{0.3\textwidth}
        \centering
        \includegraphics[scale = 0.28, trim = 4cm 0cm 5cm 0cm, clip]{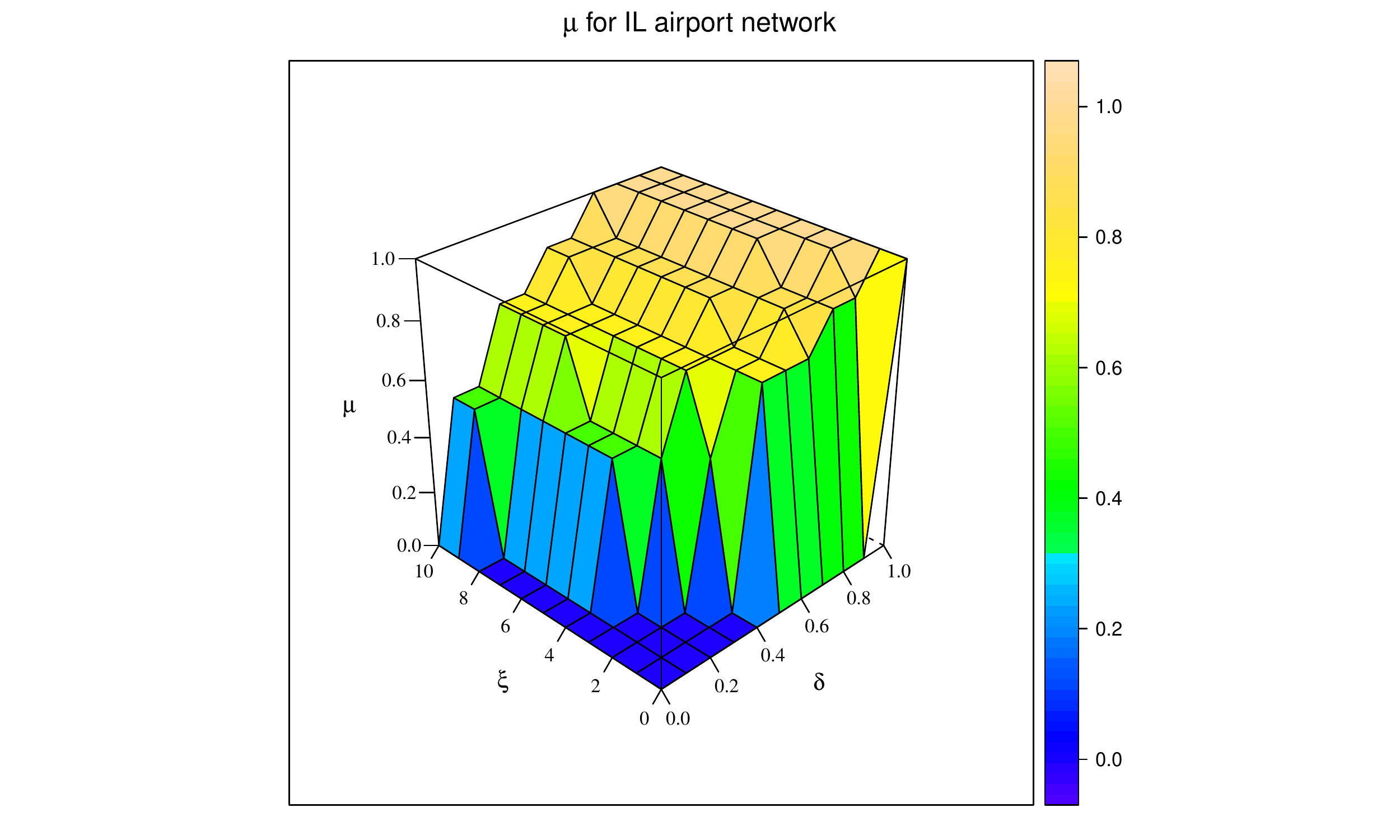}
        %\caption{$\Theta_2$}
     \end{minipage}%
     \begin{minipage}{0.3\textwidth}
        \centering
        \includegraphics[scale = 0.28, trim = 4cm 0cm 5cm 0cm, clip]{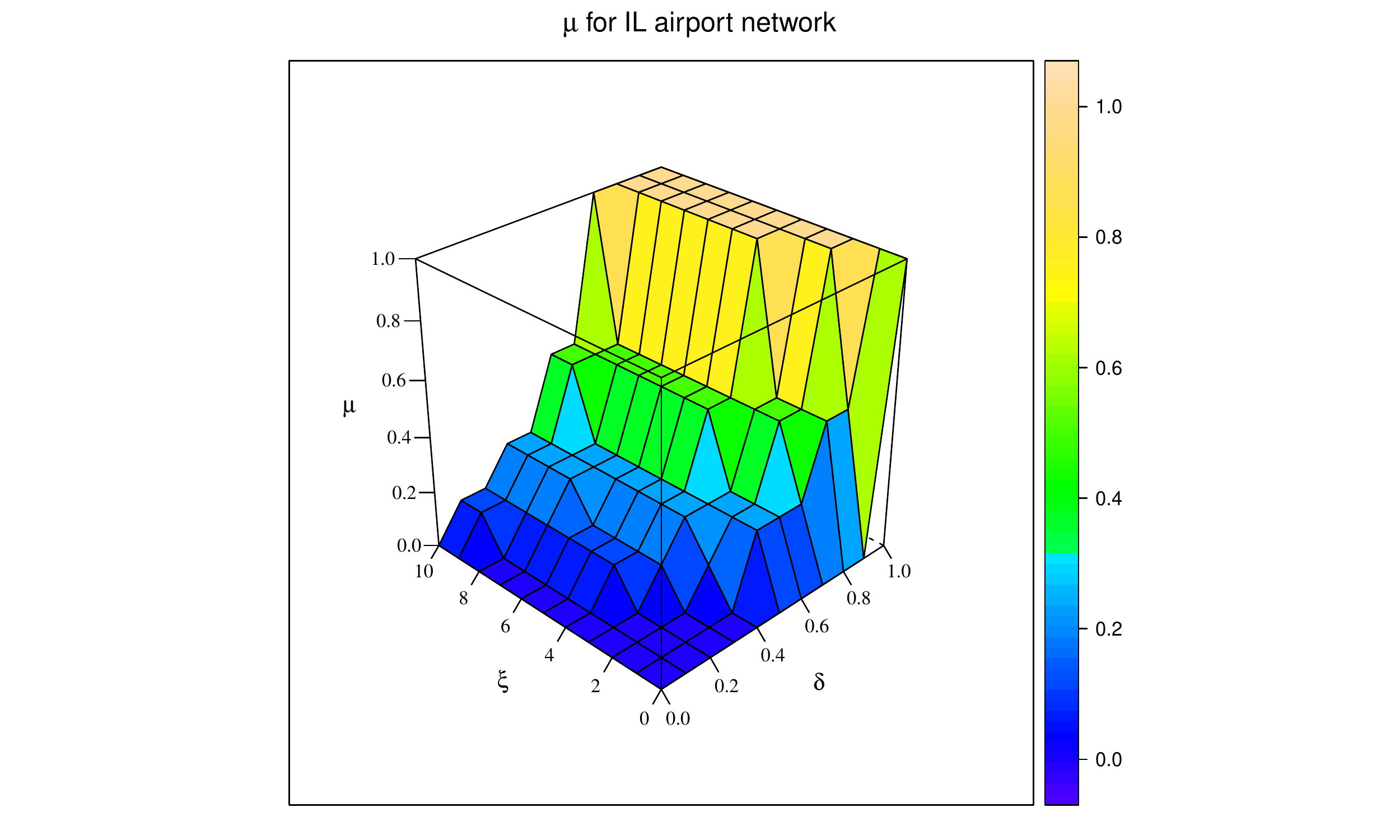}
        %\caption{$\Theta_3$}
     \end{minipage}
    \caption{$\mu_{(\Gamma,\xi)}(N)$  values in case of IL state airports for $\Gamma_2$ and $\Theta_1$ (left), $\Theta_2$ (center) and $\Theta_3$ (right).}
    \label{il_2}
\end{figure}

When we consider $\Gamma_2$ (see Figure~\ref{il_2}), the $\gamma$'s
are constant for the first $\bar{k}/2$ values, and then they start
to grow; thus, we are penalizing longest simple paths. In this case
(with $\Theta_1$), the effect of $\delta$ is more prominent than the
one of $\xi$, and $\mu_{(\Gamma,\xi)}(N)$ tends to 1 for high
values of $\delta$. Such a trend can be observed also for $\Theta_2$
and $\Theta_3$. However, $\Theta_2$ is much more similar to
$\Theta_1$ than $\Theta_3$, because its longest paths are penalized from
the $\gamma$'s also in this setting; thus, the damping effect of
the second part of $\Theta_2$ in the weight vector seems to be
hardly noticeable. The penalization of the longest path is also the
reason for the absence of an evident collapse effect for $\Theta_3$,
because shock propagation is impeded on longest paths by $\gamma$'s
and on shortest paths by $\theta$'s.

\begin{figure}[!htb]
    \centering
    \begin{minipage}{0.3\textwidth}
        \centering
        \includegraphics[scale = 0.28, trim = 4cm 0cm 5cm 0cm, clip]{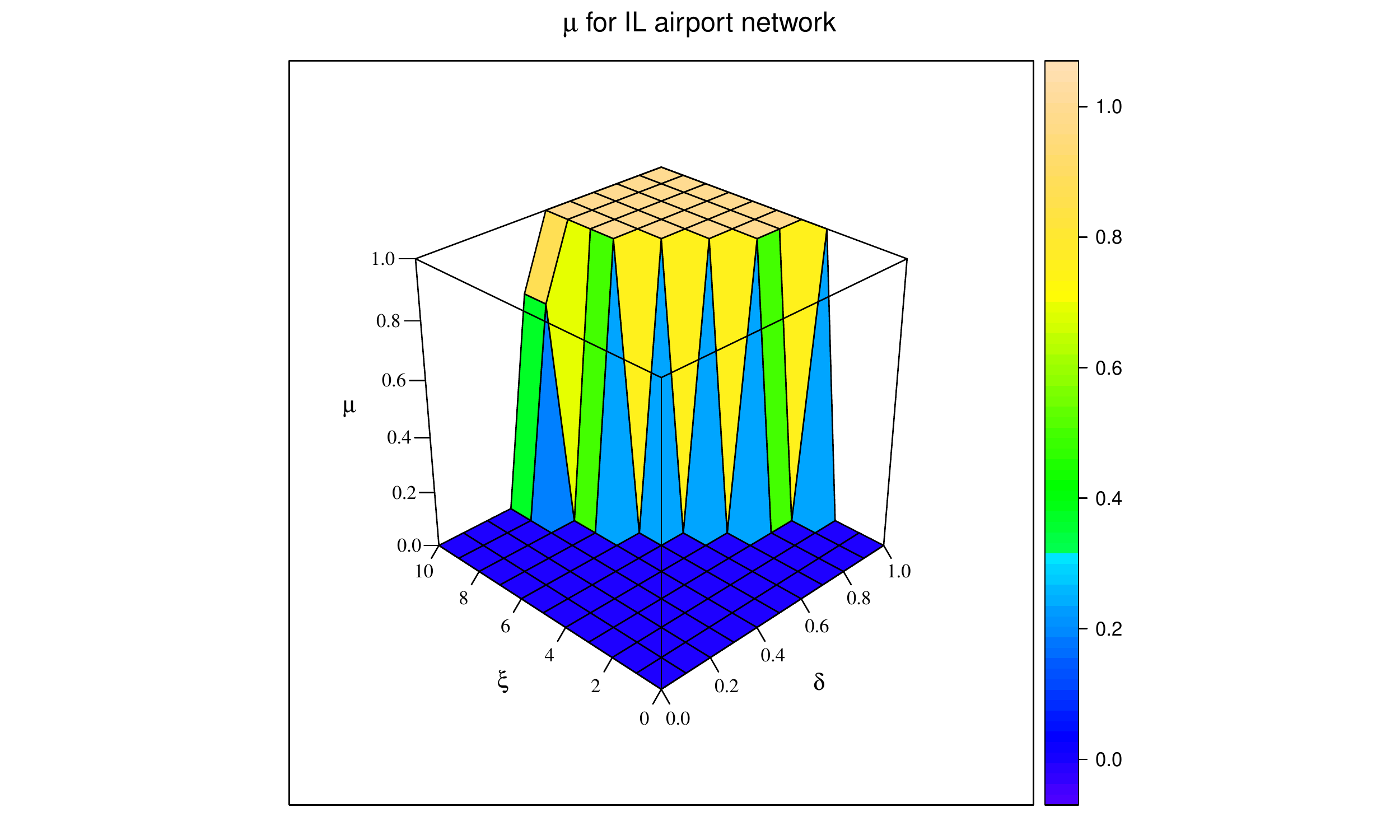}
        %\caption{$\Theta_1$}
    \end{minipage}%
     %\ \hspace{5mm} \hspace{5mm}
    \begin{minipage}{0.3\textwidth}
        \centering
        \includegraphics[scale = 0.28, trim = 4cm 0cm 5cm 0cm, clip]{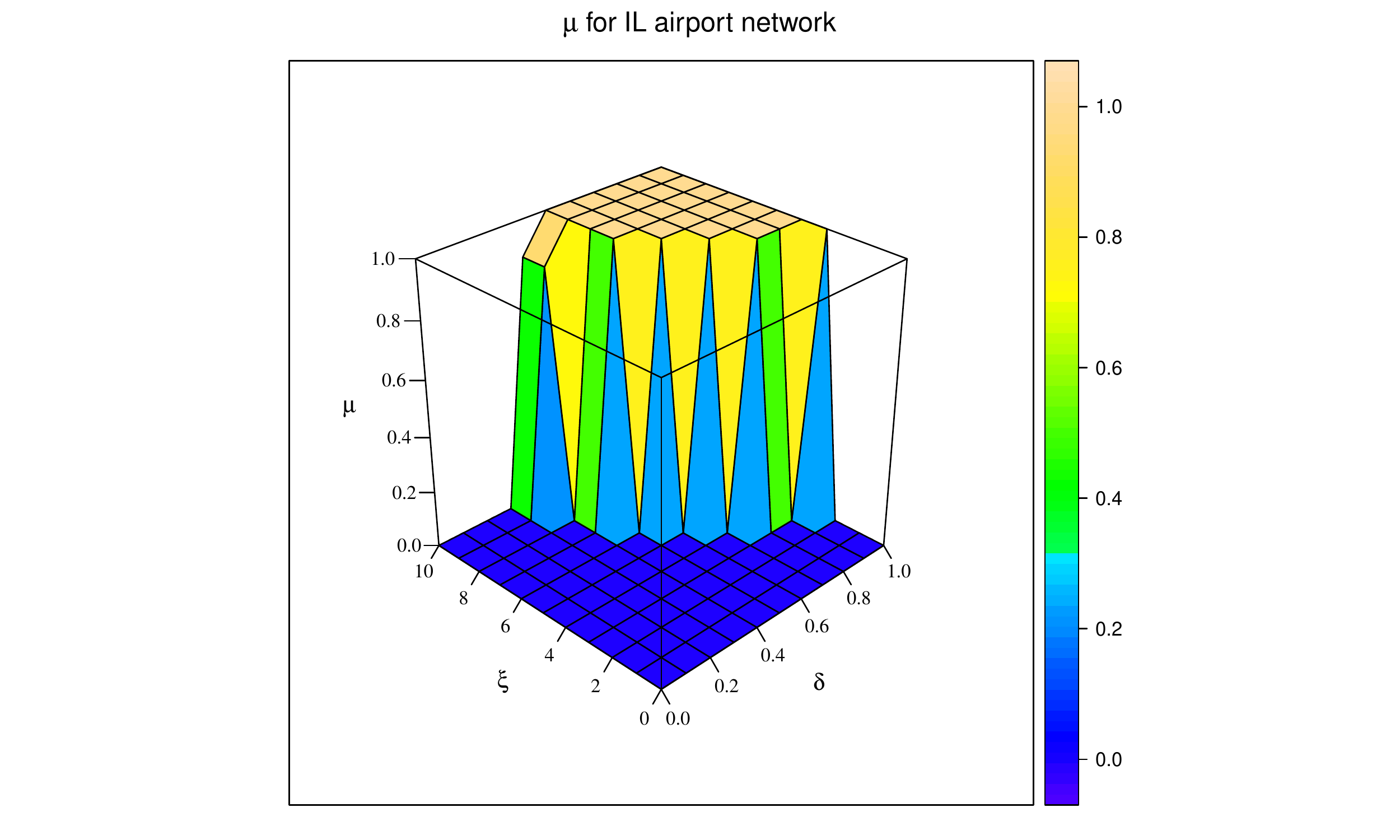}
        %\caption{$\Theta_2$}
     \end{minipage}%
     \begin{minipage}{0.3\textwidth}
        \centering
        \includegraphics[scale = 0.28, trim = 4cm 0cm 5cm 0cm, clip]{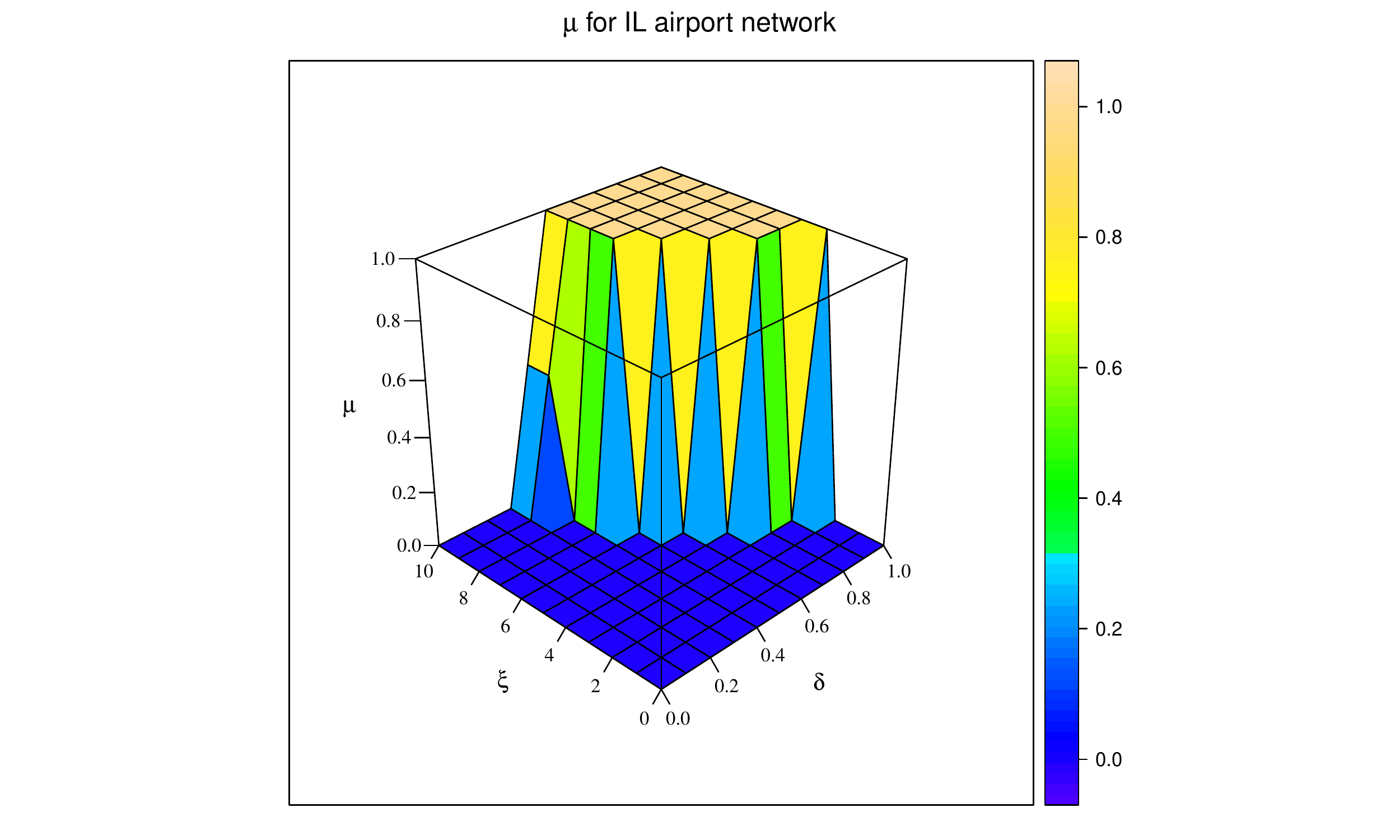}
        %\caption{$\Theta_3$}
     \end{minipage}
    \caption{$\mu_{(\Gamma,\xi)}(N)$  values in case of IL state airports for $\Gamma_3$ and $\Theta_1$ (left), $\Theta_2$ (center) and $\Theta_3$ (right).}
    \label{il_3}
\end{figure}

The third setting of $\Gamma$ is associated to the penalization of
the shortest simple paths, so that the shock does not propagate
unless it has a high value and the discount factor is high as well.
Figure~\ref{il_3} reports the results, which are similar for all
settings of $\Theta$.

%[XXX RIMUOVEREI, NON MI SEMBRA CHIARISSIMO ED E' UN CAMBIAMENTO
%DAVVERO IMPERCETTIBILE It is worth to note that when considering
%$\Theta_1$ and $\Theta_2$ we have returns of scale different from
%the case of $\Theta_3$, since in the latter the trend, for each
%$\xi$, goes from concave to convex. Therefore, when shortest paths
%are penalized, the value of $\delta$ has an importance greater than
%the magnitude of the shock $\xi$. XXXX]

\subsection{New York state network}

\begin{figure}[!htb]
    \centering
    \begin{minipage}{0.3\textwidth}
        \centering
        \includegraphics[scale = 0.28, trim = 4cm 0cm 5cm 0cm, clip]{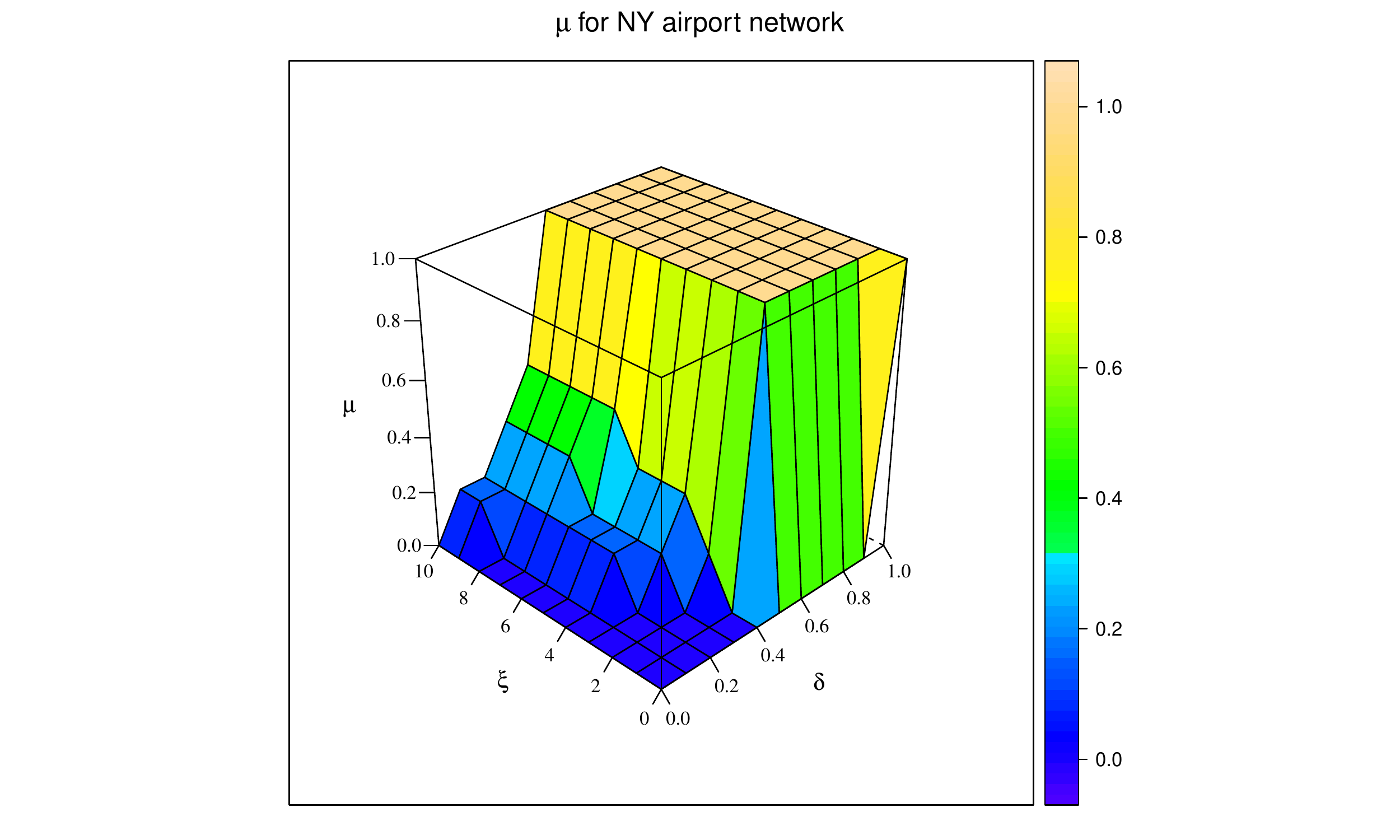}
        %\caption{$\Theta_1$}
    \end{minipage}%
     %\ \hspace{5mm} \hspace{5mm}
    \begin{minipage}{0.3\textwidth}
        \centering
        \includegraphics[scale = 0.28, trim = 4cm 0cm 5cm 0cm, clip]{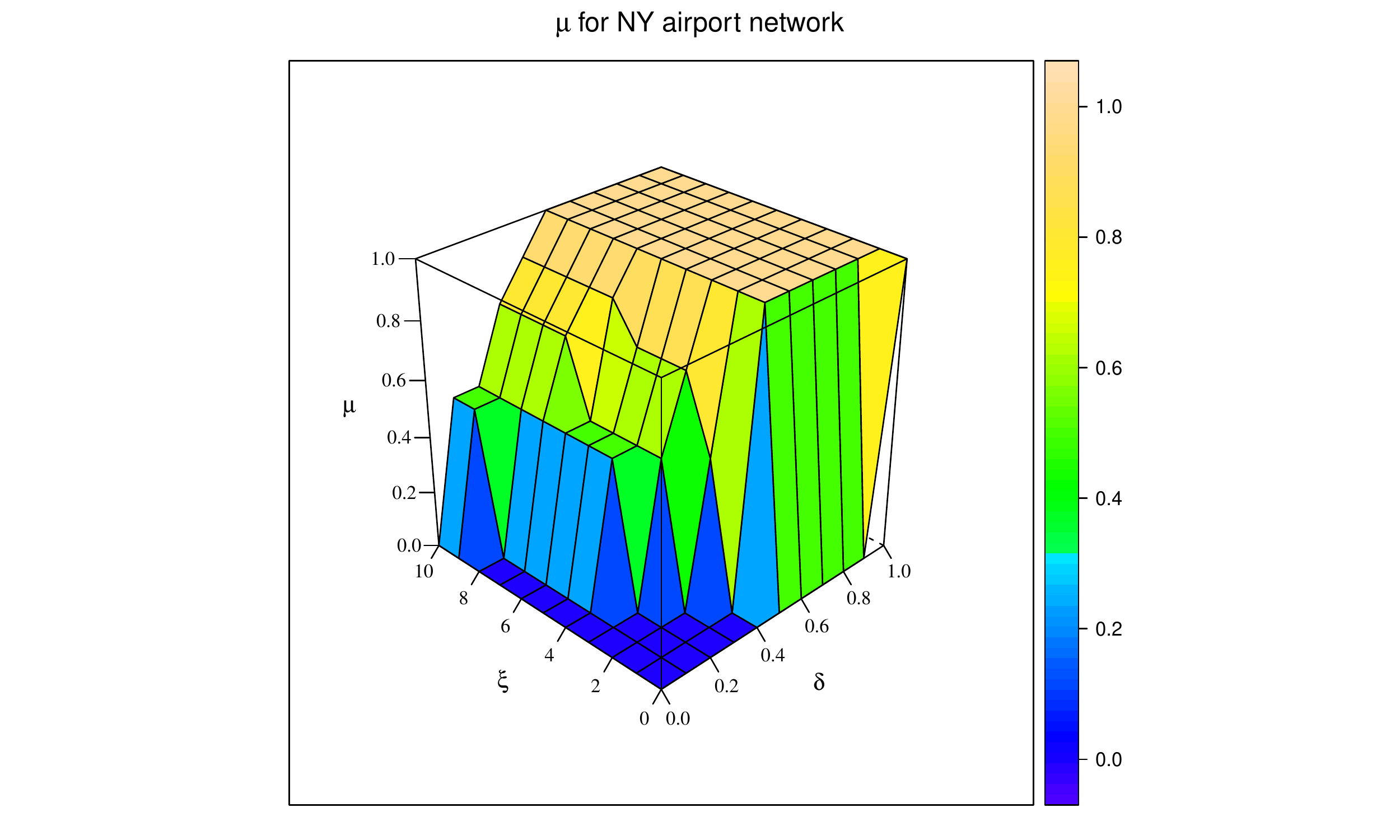}
        %\caption{$\Theta_2$}
     \end{minipage}%
     \begin{minipage}{0.3\textwidth}
        \centering
        \includegraphics[scale = 0.28, trim = 4cm 0cm 5cm 0cm, clip]{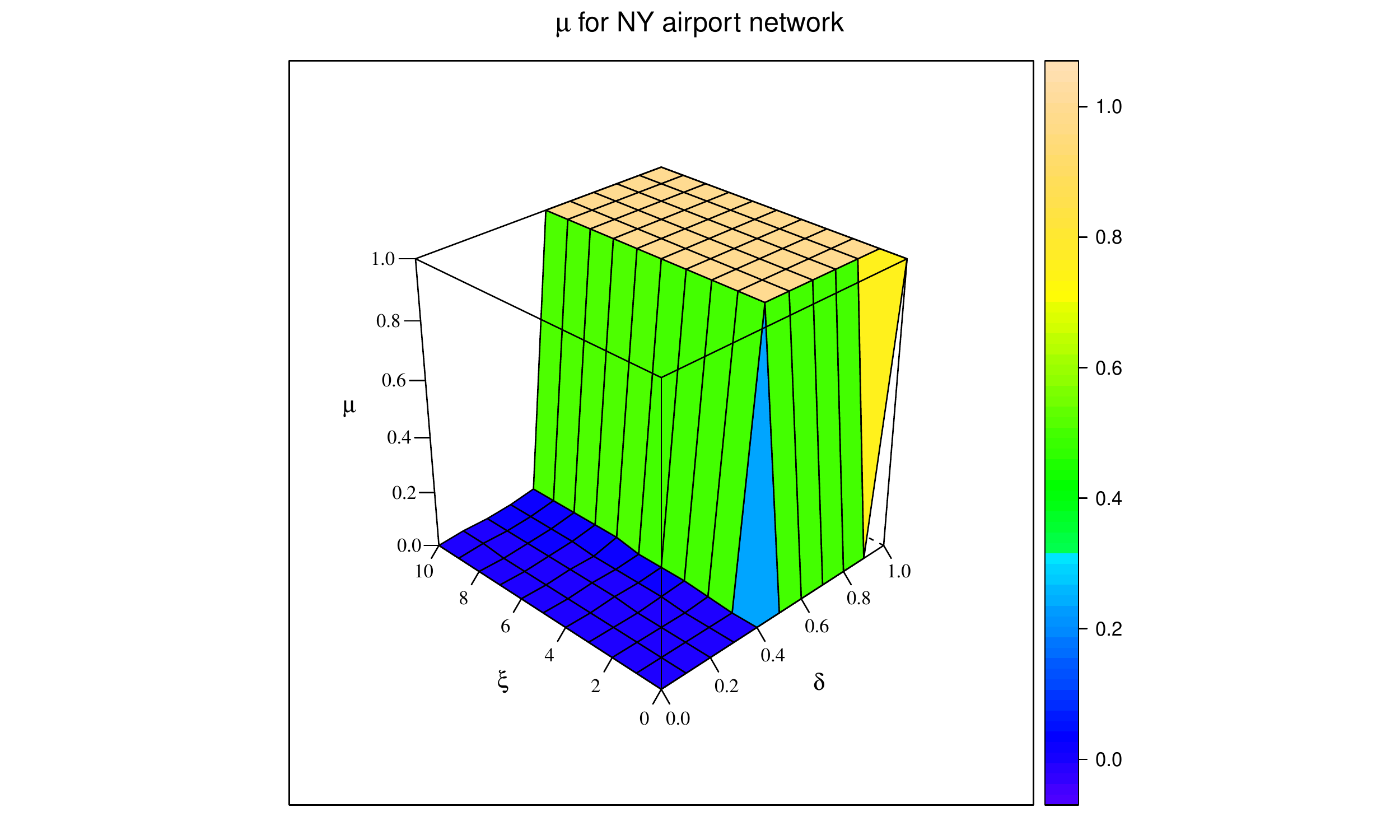}
        %\caption{$\Theta_3$}
     \end{minipage}
    \caption{$\mu_{(\Gamma,\xi)}(N)$  values in case of NY state airports for $\Gamma_1$ and $\Theta_1$ (left), $\Theta_2$ (center) and $\Theta_3$ (right).}
    \label{ny_1}
\end{figure}

In Figure~\ref{ny_1} we report the results for $\Gamma_1$ for the NY
state airport network. The behavior is very similar to the previous
network in both settings of $\Theta$ (see Figure~\ref{il_1}) but for
$\Theta_3$ the collapse effect of the resilience is very clear
for $\delta$ going above 0.4.

\begin{figure}[!htb]
    \centering
    \begin{minipage}{0.3\textwidth}
        \centering
        \includegraphics[scale = 0.28, trim = 4cm 0cm 5cm 0cm, clip]{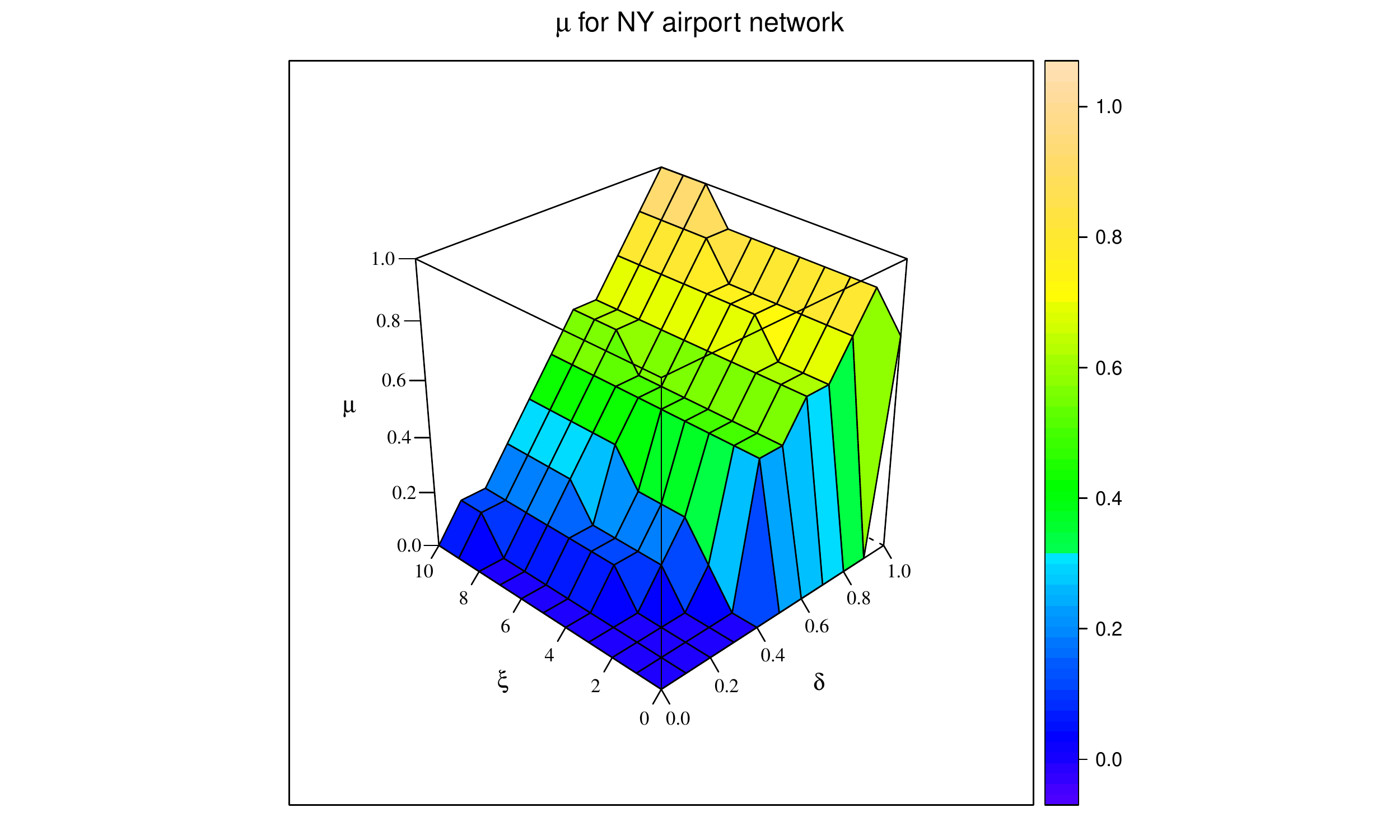}
        %\caption{$\Theta_1$}
    \end{minipage}%
     %\ \hspace{5mm} \hspace{5mm}
    \begin{minipage}{0.3\textwidth}
        \centering
        \includegraphics[scale = 0.28, trim = 4cm 0cm 5cm 0cm, clip]{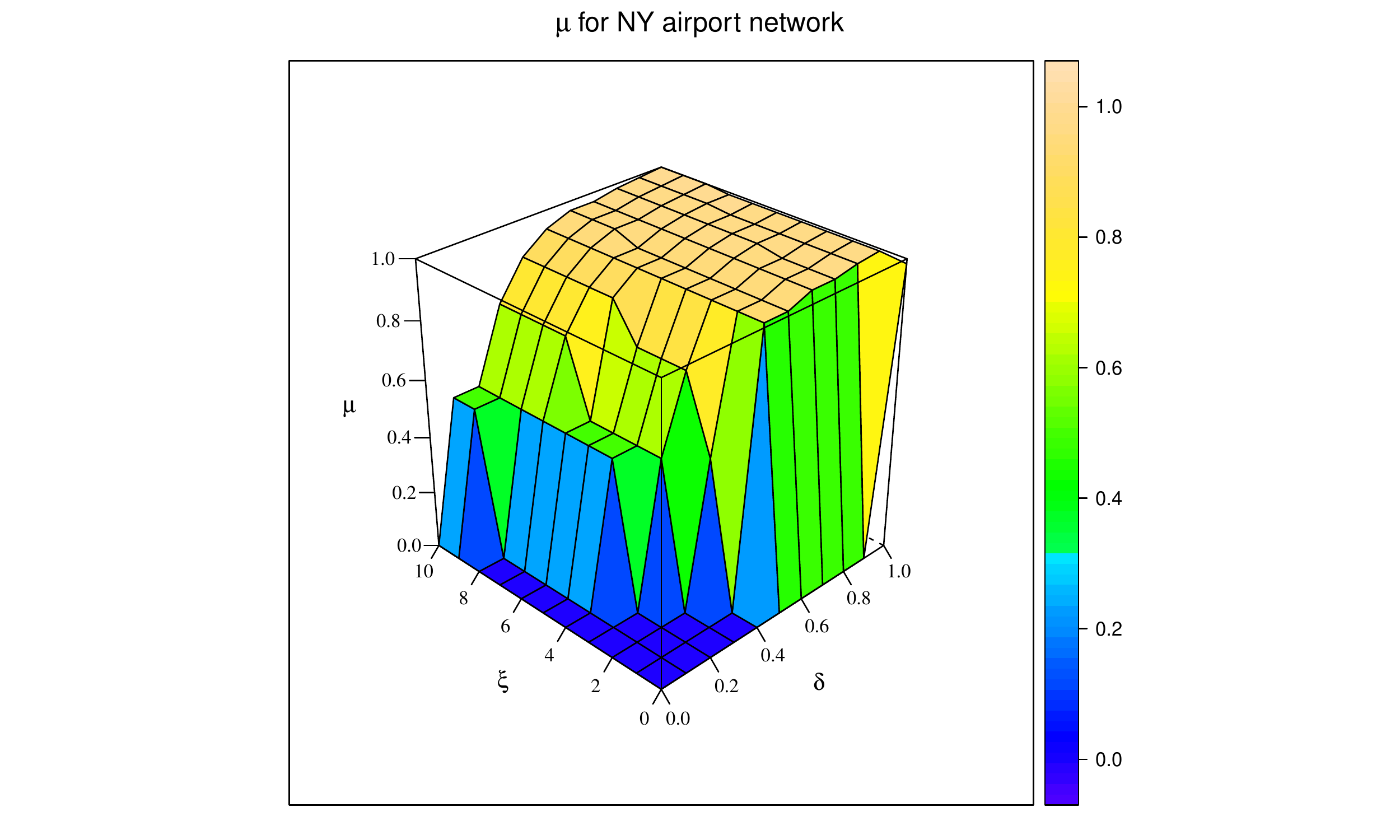}
        %\caption{$\Theta_2$}
     \end{minipage}%
     \begin{minipage}{0.3\textwidth}
        \centering
        \includegraphics[scale = 0.28, trim = 4cm 0cm 5cm 0cm, clip]{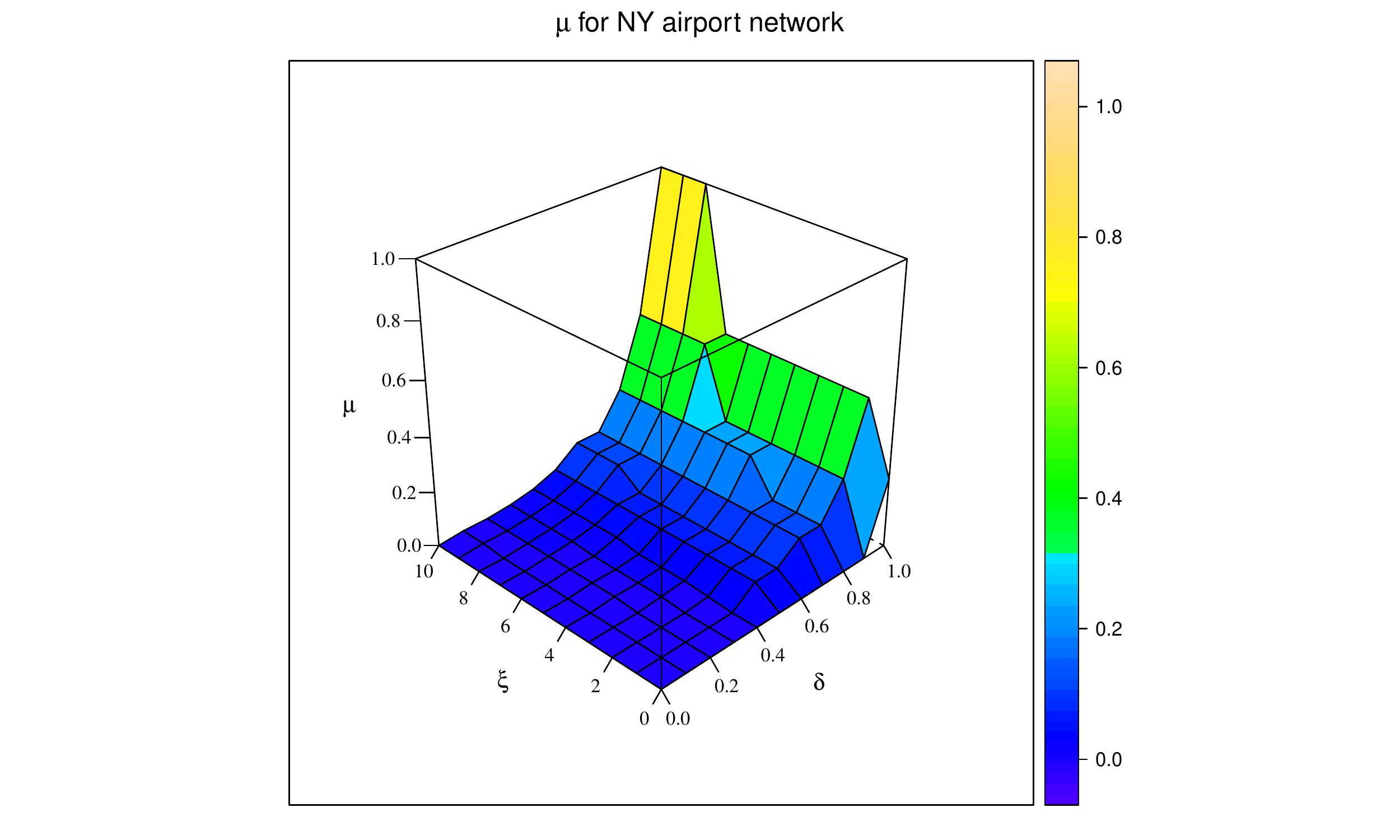}
        %\caption{$\Theta_3$}
     \end{minipage}
    \caption{$\mu_{(\Gamma,\xi)}(N)$  values in case of NY state airports for $\Gamma_2$ and $\Theta_1$ (left), $\Theta_2$ (center) and $\Theta_3$ (right).}
    \label{ny_2}
\end{figure}

When we set $\Gamma_2$ we note that for $\Theta_1$ the behavior is
similar to the IL network, even if the level of the resilience of
the NY network is below that of IL (compare Figure~\ref{ny_2} with
Figure~\ref{il_2}). When we consider $\Theta_2$, the surface is
smoothed in comparison to the same case for the IL network and it is
quite similar to the $\Gamma_1$ setting. For $\Theta_3$, the values
of $\mu_{(\Gamma,\xi)}(N)$ are equal to 1 only for high values of
$\xi$ and $\delta$. Indeed, the action of the $\gamma$'s, which
contribute to deteriorating the resilience of the networks at the
short paths level, is more effective than that of the $\theta$'s,
which reduce the relevance of the the short paths in the definition
of $\mu_{(\Gamma,\xi)}(N)$.

Comparison results between NY and IL networks can be explained by
their dimensions, being that $\bar{k}=8$ for the former and
$\bar{k}=4$ for the latter. In the same light, one should interpret
also the case of $\Gamma_3$.

Indeed, when considering the NY airports, all the values of
$\mu_{(\Gamma,\xi)}(N)$ are equal to 0 and Figures are not shown.
In this setting one has $\gamma_1>10$, which is the maximum value of
the available $\xi$, so that propagation cannot occur.

\section{Conclusive remarks}

The analysis of the resilience of networks and their attitude to
wards absorbing  external shocks is crucial for the various implications
related to reliability and functionality of real-world engineering
systems.

The aim of this paper is to propose a new measure of network
resilience based on the study of the shocks propagation along the
patterns of connections among nodes.

In our setting, shocks are assumed to have a positive size and a tendency of
not necessarily removing nodes from the network. They propagate in
an amplified or dampened way through a suitably defined discount
factor, with such a propagation proceeding under the fulfilment of a
predefined condition involving the size of the shocks and the
weights of the arcs.

The resilience measure is conceptualized as a weighted combination
of the cardinality of the sets collecting the paths with different
lengths, whereby the weights are able to assign relevance to the short
or long paths in the networks.

We test the measure on two real airport systems.
The comparison between the empirical instances suggests the incidence
of the size of the network on shock propagation, as well as on the
resilience. Indeed, for large networks, propagation suffers from
difficulty in reaching nodes that are far in respect to the others.

Results state that the resilience is strongly dependent on the
weights associated to the short paths, and highlight a prominent
role of the discount factor in determining the ability of the
network to absorb shocks. This behavior suggests that it is possible
to design resilient network, with a strong ability to absorb
external shocks, by imposing a propagation pattern on the basis of a
certain level of the discount factor.

The proposed resilience measure might then be of usefulness to the
engineers for the identification of the vulnerabilities of systems
and for designing more resilient infrastructures.

\end{document}